\documentclass[12pt]{iopart} 
\usepackage{amssymb}
\usepackage{graphicx}
\usepackage{harvard}
\usepackage[dvips]{color}

\newcommand{\tj}[6]{\left(\begin{array}{lcr}#1 & #2 & #3 \\ #4 & #5 &
      #6
                    \end{array}\right) } 

\begin{document}   
   
\jl{2}
   
\title[Near threshold rotational excitation of molecular ions by
electron-impact] {Near threshold rotational excitation of molecular
ions by electron-impact}
   
\author{A Faure$^1$, V Kokoouline$^2$, Chris H Greene$^3$ and Jonathan
Tennyson$^4$}
   
\address{$^1$ Laboratoire d'Astrophysique, UMR 5571 CNRS, Universit\'e
Joseph-Fourier, B.P. 53, 38041 Grenoble cedex 09, France}

\address{$^2$ Department of Physics, University of Central Florida,
Orlando, Florida 32816, USA}

\address{$^3$ Department of Physics and JILA, University of Colorado,
Boulder, Colorado 80309-0440, USA}

\address{$^4$ Department of Physics and Astronomy, University College
  London, Gower Street, London WC1E 6BT, UK}

\ead{afaure@obs.ujf-grenoble.fr}
   
\begin{abstract}   
  
New cross sections for the rotational excitation of H$_3^+$ by
electrons are calculated {\it ab initio} at low impact energies. The
validity of the adiabatic-nuclei-rotation (ANR) approximation,
combined with $R$-matrix wavefunctions, is assessed by comparison with
rovibrational quantum defect theory calculations based on the
treatment of Kokoouline and Greene ({\it Phys. Rev. A } {\bf 68}
012703 2003). Pure ANR excitation cross sections are shown to be
accurate down to threshold, except in the presence of large
oscillating Rydberg resonances. These resonances occur for transitions
with $\Delta J=1$ and are caused by closed channel effects. A simple
analytic formula is derived for averaging the rotational probabilities
over such resonances in a 3-channel problem. In accord with the Wigner
law for an attractive Coulomb field, rotational excitation cross
sections are shown to be large and finite at threshold, with a
significant but moderate contribution from closed channels.

\end{abstract}

\section{Introduction}
   
Rotational excitation of positive molecular ions is a major mechanism
by which slow electrons lose energy in partially ionized molecular
gas. The efficiency of this process is a significant parameter in
various applications, for example the interpretation of dissociative
recombination experiments \cite{lammich03,kokoouline03} or the
determination of the density and temperature conditions in the diffuse
interstellar medium \cite{lim99,faure06}. A good knowledge of
rotational (de)excitation cross sections is therefore required for a
variety of molecular ions and over a wide range of collisional
energies and rotational levels. In contrast to neutral molecules,
however, little attention has been given to the electron-impact
excitation of molecular ions. These are indeed more difficult to study
experimentally and only a few theoretical treatments have been
reported so far. The reference method for computing electron-impact
excitation cross sections has been the Coulomb-Born (CB) approximation
\cite{chu74,chu75,dickinson77,neufeld89}. This approach assumes that
the collisional cross sections are entirely dominated by long-range
interactions. The CB theory thus predicts that transitions with
$\Delta J$=1 and 2 only are allowed for dipole and quadrupole
interactions, respectively. Recent {\it ab initio} $R$-matrix studies
on several astronomically important molecular ions have shown,
however, that this prediction is incorrect and that the inclusion of
short-range interactions is crucial, particularly for ions with small
(or zero) dipole or transitions with $\Delta J>1$ (see
\citeasnoun{rabadan98}, \citeasnoun{faure02c} and references therein).

These $R$-matrix studies were, however, hampered by the use of the
adiabatic-nuclei-rotation (ANR) approximation. The ANR theory is
expected to become invalid close to a rotational threshold because it
neglects the rotational Hamiltonian (see, e.g.,
\citeasnoun{lane80}). This leads to an ambiguous interpretation of the
ANR energy because the electron energy in the exit channel of a
near-threshold inelastic collision is very small while the assumption
of rotational degeneracy sets this energy equal to that of the
entrance channel. As a result, if the ANR energy is interpreted (as
usual) as the initial kinetic energy, then excitation cross sections
are non-zero at and below threshold. In previous $R$-matrix works,
this artefact was corrected by multiplying the ANR cross sections by
the kinematic ratio $k'/k$, where $k$ ($k'$) is the initial (final)
momentum of the electron. This simple method, which forces the cross
sections to zero at threshold, is widely used in the case of neutral
molecules \cite{morrison88}. The resulting cross sections, however, do
not obey the proper threshold law, when the molecular target has a
nonzero charge. In the present work, near threshold rotational
excitation of H$_3^+$ by electron-impact is investigated by comparing
ANR cross sections with independent calculations based on the quantum
defect theory and rotational-frame-transformation (MQDT-RFT) approach
\cite{fano70,chang72,child90,kokoouline03} To our knowledge, this is
the first quantitative assessment of the validity of the adiabatic
approximation for electron-impact rotational excitation of a charged
target molecule. The theoretical treatments are introduced in the next
section. Comparisons of cross sections obtained at different levels of
theory are presented in section~3. Conclusions are summarized in
section~4.

\section{Theory}

We consider the following process between an electron and a
symmetric-top molecular ion:
\begin{equation}
e^-(E_k) +ION(JKM)\rightarrow e^-(E'_k) +ION(J'K'M'),
\end{equation}
where $J$ is the ionic rotational angular momentum, $K$ and $M$ its
projections along the body-fixed (BF) and the space-fixed (SF) axes,
respectively, and $E_k$ ($E'_k$) is the initial (final) kinetic energy
of the electron. The relation between $E_k$ and $E'_k$ is:
\begin{equation}
E'_k=E_k+(E_{JK}-E_{J'K'}),
\end{equation}
where the energy of the state (JK) is given in the rigid rotor approximation by:
\begin{equation}
E_{JK}=BJ(J+1)+(A-B)K^2,
\end{equation}
with $A$ and $B$ being the rotational constants. It should be noted
that vibrational energy splittings of H$_3^+$ are much larger than
rotational splittings, with the first excited vibrational level
$\{01^1\}$ being 0.3~eV above the ground vibrational level
$\{00^0\}$. In the following, we will assume that the initial and
final vibrational state of the ion is $\{00^0\}$ and that the incident
electron energy is less than 0.3~eV.

\subsection{ANR method}
 
We follow the implementation of the ANR theory for symmetric-top
molecular ions as presented in \citeasnoun{faure02c}. The key
quantities in this approach are the $T$-matrix elements which are
produced in the BF frame of reference and are then transformed through
the conversion of the BF asymptotic scattering amplitude to the SF
frame. In this approach, the rotational Hamiltonian is neglected and
there is an ambiguity in defining the BF asymptotic electron energy,
$E_{bf}$, which can be taken equal to the kinetic energy in the
entrance ($E_k$) or exit ($E'_k$) channels, or even to some other
values (see \citeasnoun{morrison88} and references for neutral targets
therein). The usual choice, however, is to take $E_{bf}$ as the
initial kinetic energy, $E_k$. The integral ANR rotationally inelastic
cross section for a symmetric-top molecule is then given by
\cite{faure02c}:
\begin{equation}
\hspace{-1.5cm} \sigma^{\rm ANR}_{JK\rightarrow J'K'}(E_k) =
\frac{(2J'+1)\pi}{2E_k}\sum_{jm_j}(2j+1 ) \times
\tj{J}{J'}{j}{K}{-K'}{m_j}^2\sum_{ll'}|M_{ll'}^{jm_j}(E_k)|^2
\label{sig1}
\end{equation}
with
\begin{equation}
M_{ll'}^{jm_j}(E_k)=\sum_{mm'hh'p\mu}
\bar{b}_{lhm}^{p\mu}\tj{l}{l'}{j}{-m}{m'}{m_j}\bar{b}_{l'h'm'}^{p\mu}T_{lh,l'h'}^{p\mu}(E_k)
\end{equation}
where $l$ ($l'$) is the initial (final) electron orbital momentum,
$T_{lh,l'h'}^{p\mu}$ are the BF $T$-matrix elements belonging to the
$\mu$th component of the $p$th irreducible representation (IR) of the
molecular point group, with $h$ distinguishing between different
elements with the same ($p\mu l$). The $\bar{b}$ coefficients are
discussed, for example, in \citeasnoun{gianturco86}. The {\it
T}-matrix is as usually defined in terms of {\it S}-matrix as
\begin{equation}
\bi{T}=\bf{1}-\bf{S}
\end{equation}
and the cross section can be expressed in terms of the SF {\it
S}-matrix by the familiar formula:
\begin{equation}
\sigma^{\rm ANR}_{JK\rightarrow
J'K'}(E_k)=\frac{\pi}{2E_k(2J+1)}\sum_{J_{\rm tot}ll'}(2J_{\rm
tot}+1)|\delta_{JJ'}\delta_{KK'}\delta_{ll'}-S^{J_{\rm
tot}}_{JKl,J'K'l'}(E_k)|^2,
\label{sig2}
\end{equation}
where $\bi{J_{\rm tot}}=\bi{J}+\bi{l}$ is the total angular
momentum. It should be noted that the ambiguous choice $E_{bf}=E_k$
leads to excitation cross section that are non-zero below threshold. A
common method for forcing the cross sections to zero at threshold is
to multiply these by the ratio of the final and initial momentum of
the electron, $k'/k$ \cite{morrison88}. This was used in particular in
previous $R$-matrix works \cite{faure02c}. However, the ANR $T$-matrix
elements still do not exhibit the proper dependence on $k'$ in this
limit, for an ionic target. Alternatives such as $E_{bf}=E_k'$ are
possible but still should not give the correct threshold laws for the
$T$-matrix elements. Furthermore, the principle of detailed balance is
not rigorously satisfied near threshold. This will be further
discussed in section~2.3.

The $e^-$+H$_3^+$ BF $T$-matrices were taken from the $R$-matrix
calculations of \citeasnoun{faure02b}. The geometry of the ion was
frozen at its equilibrium position corresponding to an equilateral
triangle with D$_{3h}$ symmetry. The $e^-$+H$_3^+$ scattering model
was constructed using the four lowest target electronic states and the
continuum orbitals were represented using the GTO basis set given by
\citeasnoun{faure02a} which include all angular momentum up to $l_{\rm
max}=4$ and is optimized to span energies below 5~Ryd. As discussed in
\citeasnoun{faure02c}, rotationally inelastic cross sections were
found to be entirely dominated by low partial waves, with more than
90~\% arising from $l=1$, and ANR cross sections were obviously
unaltered by augmenting them with quadrupolar CB calculations for
partial waves with $l>4$. Collisional selection rules are detailed in
\citeasnoun{faure02c}.

\subsection{MQDT-RFT method}

We follow the implementation of the MQDT-RFT theory as presented by
Kokoouline and Greene (2003, 2004). The $e^-$+H$_3^+$ scattering model
is based on {\it ab initio} H$_3^+$ and H$_3$ potentials independent
from the $R$-matrix model presented above. In this framework,
originally developed for the dissociative recombination of H$_3^+$,
first, one constructs the energy-independent $S^{BF}$-matrix in a
basis better adapted for BF (see Kokoouline and Greene (2003, 2004)
and references therein).  The $S^{BF}$-matrix depends on the three
internuclear coordinates $\cal Q$ and the projection $\Lambda$ of the
electronic orbital momentum on the molecular axis. Thus, we define the
elements of the scattering matrix as
\begin{equation}
\label{eq:S-BF}
\langle {\cal Q};\Lambda \big| \hat S\big|{\cal Q}';\Lambda '\rangle =S^{BF}_{\Lambda;\Lambda'}({\cal Q})\delta({\cal Q},{\cal Q}').
\end{equation}
The $S^{BF}$-matrix in this representation is diagonal with respect to
the rotational quantum numbers $J_{\rm tot},\,K_{\rm tot}$ and $M_{\rm
tot}$ of the whole molecule and the continuous coordinate $\cal Q$. To
completely define basis functions $|b\rangle$ in the BF, in addition
to $\Lambda$ and $\cal Q$, the rotational quantum numbers $J_{\rm
tot},\,K_{\rm tot}$ and $M_{\rm tot}$ must be also specified; for
brevity they are omitted in the above equation. For the matrix element
$S^{BF}_{\Lambda;\Lambda'}({\cal Q})$ we will also use the notation $
\langle b|\hat S|b'\rangle$. The next step in the MQDT-RFT is to
obtain the $S$-matrix in the basis $|s\rangle$ corresponding to the
laboratory or SF coordinate system.  In this basis, the good quantum
numbers are the vibrational quantum numbers (for example,
\{$v_1,v_2^{l_2}$\}, if the normal mode approximation is used) and
rotational quantum numbers $J_{\rm tot}, J, K, M,l, m$ defined
above. In the following, we will not specify any other conserved
quantum numbers that are the same in the both bases, such as the total
nuclear spin and the irreducible representation of the total
wavefunction. The SF $S$-matrix is determined by carrying out a
unitary transformation from the BF basis to the SF basis:
\begin{eqnarray}
\label{eq:RVFT}
S_{s;s'}=\sum_{b,b'} \langle s|b\rangle  \langle b|\hat S|b'\rangle \langle b'|s'\rangle\,,
\end{eqnarray}
where the summation indicates a sum over discrete indices and an
integration over the continuous coordinates $\cal Q$. The explicit
form of the the unitary transformation matrix elements $\langle
s|b\rangle$ is given in Eq.(27) of \citeasnoun{kokoouline04}.

The transformation of Eq. \ref{eq:RVFT} is truly unitary only if the
vibrational basis in the SF representation is complete. In
calculations, one always adopts a finite vibrational basis set that is
not complete. It should be sufficiently large to represent properly
the vibrational states contributing to the process of interest. In
practice the vibrational basis set is calculated within a finite
volume, and one can impose outgoing-wave (Siegert) boundary conditions
at its surface to account for the fact that some of the incident
electron flux can be diverted through the collision into dissociative
channels, as in \citeasnoun{hamilton02} and \citeasnoun{kokoouline03}.
The procedure of transformation from the BF to SF representations
described above is called the rovibrational frame transformation in
MQDT. It accounts for the coupling between electronic, rotational and
vibrational motion of the molecule. In contrast to the MQDT-RFT
approach, the rotational dynamics of the system is not included in the
ANR.

The unitary transformation of the $S$-matrix from the body frame to
the laboratory frame, as described in the preceding paragraph, is
similar in spirit to that described by Chase (1956,1957). But one
important physical difference is that, whereas the Chase approximation
was intended to be applied only in energetically open collision
channels, in the MQDT-RFT the transformed scattering matrix $S$
includes (weakly) closed channels as well. When the incident electron
collides with a singly-charged molecular ion of size $r_0$, its
kinetic energy is increased by $1/r_0$, and consequently scattering
into weakly-closed channels is often an important class of pathways,
even for entrance or exit scattering processes that occur right at
threshold with zero asymptotic kinetic energy.  Every such threshold
has an infinite number of Rydberg levels converging to it.  For this
reason, the transformed $S$ does not yet represent the physical
scattering matrix \cite{aymar96}. In fact, it represents the actual
scattering matrix $S^{phys}$ only in energy ranges where {\it all} of
the channels $|s\rangle$ are open for electron escape, i.e. where the
total energy of the system is higher than the energy of the highest
relevant ionization channel $|s\rangle$ threshold. When at least one
channel is closed, the physical scattering matrix $S^{phys}$ is
obtained from $S$ using the standard MQDT channel-elimination formula
(see Eq. (2.50) in \citeasnoun{aymar96} or Eq. (38) in
\citeasnoun{kokoouline04} ).

In the H$_3^+$ DR studies carried out by Kokoouline and Greene (2003,
2004), the treatment has been limited to the dominant $p$-wave
component ($l$=1) of the incident electron wavefunction because the
$p$-wave has the largest contribution into the DR cross-section. This
partial wave also has the largest effect on the rotational excitation
probabilities considered in the present study. When all channels
$|s\rangle$ are open for electron escape, the scattering matrix is
energy-independent. (This is within the frequently-adopted
approximation of energy-independent quantum defects, which can be
improved upon when necessary, as has been discussed, e.g., by
\citeasnoun{gao90}.) When there are closed channels that are coupled
to open channels, the autoionizing Rydberg series associated with the
closed ionic channels introduce a strong energy dependence into the
physical scattering matrix. Consequently, the resulting cross-sections
reflect these autoionizing resonance features that are strongly
dependent on the energy. Examples of such energy dependences that
arise in electron collisions with H$_3^+$ are demonstrated below.
These rotational autoionizing resonances have been thoroughly studied
in both experimental (\citeasnoun{bordas91}) and theoretical
(\citeasnoun{stephens95}, \citeasnoun{kokoouline04}) studies of H$_3$
photoabsorption processes.

MQDT-RFT calculations including the complete rovibrational frame
transformation and the Jahn-Teller effect have shown that the
probability of rotational excitation $|\bi{S}^2|$ is weakly
energy-dependent in the region between the $\{00^0\}$ and $\{01^1\}$
vibrational levels \cite{kokoouline03}. The Rydberg resonances present
in the rotational-excitation spectrum have in general small widths.
However, for transitions involving symmetries with more than two
rotational channels, the rotationally inelastic probabilities are
actually nearly energy-independent only at electron energies above the
highest channel (all coupled rotational levels are open).  For a
3-channel problem, there are thus an infinite number of purely
rotational Rydberg resonances at electron energies between the second
and third channel thresholds. This is illustrated in Fig.~1 for the
transition (1, 1)$\rightarrow$(2, 1) at $J_{\rm tot}=2$ (there is no
$p$-wave Rydberg series attached to (3, 1) at $J_{\rm tot} =1$). Note
that for the reverse process (2, 1)$\rightarrow$ (1, 1), all curves in
Fig.~1 would be simply shifted towards zero electron-impact energy, in
accordance with the detailed balance principle (see section~2.3).

\begin{figure}
\begin{center}
  {\resizebox{150mm}{!}{
      \includegraphics*{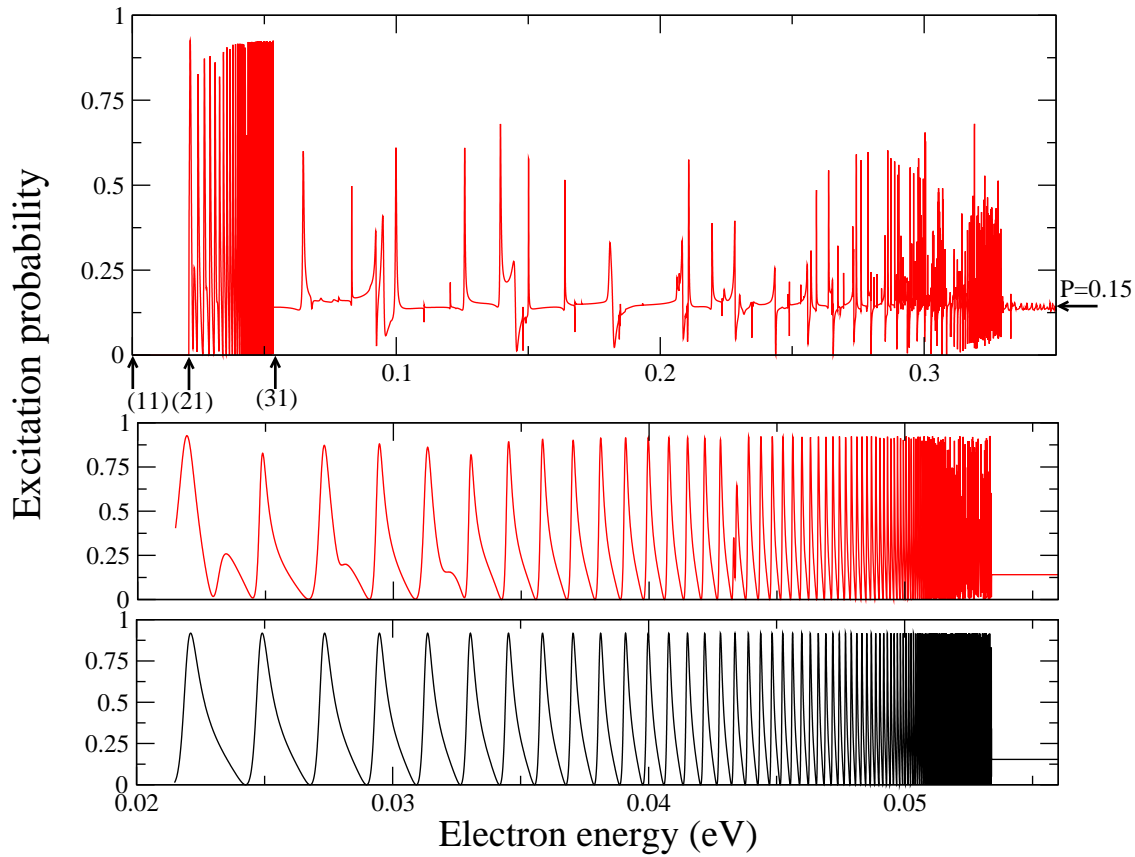}}}
\caption[] {Inelastic rotational excitation probability from the $(1,
  1)$ state of H$_3^+$ to the $(2, 1)$ state as a function of
  electron-impact energy. The probability is approximately
  energy-independent at the value 0.15 just above the (1,
  1)$\rightarrow$ (3,1) rotational threshold. A pure MQDT-RFT
  calculation is presented at the bottom, while the upper two frames
  demonstrate the prediction from a fuller MQDT calculation that
  includes rovibrational autoionizing states associated with higher
  closed channel thresholds. The arrows below the upper panel
  indicates energies of the (1,1), (2,1), and (3,1) rotational states
  of the ion.}
\end{center}
\end{figure} 

It can be seen in Fig.~1 that the pure MQDT-RFT calculation (bottom
panel) is in very good agreement with the much more expensive
rovibrational MQDT treatment (top and middle panels). In the top
panel, we observe the complicated structure introduced by the
inclusion of the vibrational motion at energies above the (3, 1)
threshold. In this energy range, there are many narrow resonances
corresponding to vibrationally excited states of bound H$_3$ Rydberg
states. If one considers the integral over the resonance region,
however, these resonances do not change the probability
significantly. Thus, as previously observed by \citeasnoun{rabadan98}
for diatomic ions, a detailed treatment of vibrational motion is
unnecessary to obtain reliable thermally-averaged rotational
excitation rate coefficients of the type normally needed in
astrophysical applications. On the other hand, in the region between
the (2, 1) and (3, 1) channels, Rydberg resonances are found to
increase the rotational probability by about a factor of 2 on average,
i.e., the value of 0.15 just above the (3, 1) channel increases
discontinuously to an averaged value of 0.29. Note that higher closed
channels involving partial waves with $l>1$ are negligible. Physically
it means that incident electrons in partial waves with $l>1$, for the
energy range considered here, can not approach the ion closely enough
to be able to excite it. From the MQDT point of view, it means that
the quantum defects for molecular states with $l>1$ are close to
zero. Indeed, calculated cross sections for transitions with $\Delta
J>2$ or $\Delta K \ne 0$ are 3 to 4 orders of magnitude smaller than
those with $\Delta J=1, 2$ and $\Delta K=0$
\cite{faure02c}. Resonances due to these low probability transitions
do not therefore contribute significantly to the cross section. In the
following, we will only consider 2- and 3-channel problems
(i.e. transitions with $\Delta J=1$ and 2) and two possible energy
ranges: above and below the highest channel threshold.

\subsubsection{Above the highest channel threshold}

Since we have restricted the frame transformation analysis in this
paper to the purely rotational FT, the ionization channels in the
energy-independent scattering matrix $S_{s;s'}$ differ only in their
rotational quantum numbers. The vibrational state is the same for all
the channels, namely, the ground vibrational state of the
ion. Therefore, the integral over the vibrational coordinates $\cal Q$
in Eq. (\ref{eq:RVFT}) (or in Eq. (27) of \citeasnoun{kokoouline04})
is the same for all the matrix elements $S_{s;s'}$. Although the BF
scattering matrix is not diagonal at all geometries in the BF
rotational quantum number $\Lambda$, the integral over the vibrational
coordinates $\cal Q$ is not zero only when $\Lambda=\Lambda'$. This is
because the non-diagonal elements of the BF matrix describe
Jahn-Teller physics, which effectively couples only vibrational states
belonging to different irreducible representations. Therefore, the
integrals over the vibrational coordinates of the ground vibrational
state can be viewed as an effective diagonal $3\times 3$ or $2\times
2$ scattering matrix with elements $\rm e^{2\pi \rmi
\bar{\bi\mu}_{\Lambda}}$, where $\bar{\bi\mu}_{\Lambda}$ represent the
quantum defects averaged over the vibrational state. They are very
close to the H$_3$ Born-Oppenheimer quantum defects at the minimum of
the H$_3^+$ potential surface.  The SF scattering matrix elements
become in this approximation (Eq. (43) of \citeasnoun{kokoouline03}):
%\begin{equation}
%S_{J, K; J', K'}^{(J_{\rm tot}, K_{\rm tot}, l)} = \sum_{\Lambda}C_{l,
%  -\Lambda; J_{\rm tot}, K_{\rm tot}}^{J', K'}\rme^{2\pi \rmi
%  \bar{\mu}_{\Lambda}}C_{l, -\Lambda; J_{\rm tot}, K_{\rm tot}}^{J,
%  K}.
%\end{equation}
\begin{eqnarray}
S_{J, K; J', K'}^{(J_{\rm tot}, K_{\rm tot}, l)} = \sum_{\Lambda} (-1)^{K+K'}\sqrt{(2J+1)(2J'+1)}\rm e^{2\pi \rmi
  \bar{\mu}_{\Lambda}} \times\\
\times\tj{l}{J_{\rm tot}}{J'}{-\Lambda}{K_{\rm tot}}{-K'} \tj{l}{J_{\rm tot}}{J}{-\Lambda}{K_{\rm tot}}{-K}.
\end{eqnarray}
%In the above equation  $C_{l, -\Lambda; J_{\rm tot}, K_{\rm tot}}^{J,K}$ 
%are Clebsch-Gordan coefficients, 
%$K_{\rm tot}$ is the projection
%of $J_{\rm tot}$ along the body-fixed (BF) axis and 
%$ \bar{\mu}_{\Lambda}$ are quantum defects with. 
In all calculations below we have used $\bar{\mu}_{0}$=0.05 and
$\bar{\mu}_{\pm 1}$=0.39 for the $p$-wave quantum defects, as in
\citeasnoun{kokoouline03}. The above SF scattering matrix is energy
independent and describes the electron-ion scattering only for
energies above the highest ionization channel. The corresponding
rotational probabilities for a set of symmetries and nuclear spins can
be found in Table~VI of \citeasnoun{kokoouline03}. Finally, the
rotationally inelastic cross section is obtained as:
\begin{equation}
\label{eq:S3_smooth}
\sigma^{\rm RFT}_{JK\rightarrow J'K'}(E_k) =
\frac{\pi}{2E_k(2J+1)}\sum_{J_{\rm tot}K_{\rm tot}}(2J_{\rm
tot}+1)|\delta_{JJ'}\delta_{KK'}-S_{J, K; J', K'}^{(J_{\rm tot},
K_{\rm tot}, l)}|^2,
\end{equation}

\subsubsection{Between the 2$^{\rm nd}$ and 3$^{\rm rd}$ channel}

The physical scattering matrix $S^{phys}$ appropriate for this energy
range is obtained from the matrix of Eq. (\ref{eq:S3_smooth}) by the
usual MQDT closed-channel elimination procedure \cite{aymar96}. To
simplify the formulae we omit here all indexes and superscripts other
than 1,2, and 3. If channels 1 and 2 are open and channel 3 is closed
the amplitude of the excitation $1\to 2$ is given by

\begin{eqnarray}
S^{phys}_{2;1}(E)=S_{2;1}-\frac{S_{2;3}S_{3;1}}{S_{3;3}-{\rm e}^{-2\rmi\beta(E)}} \\
{\rm with}\, \beta(E)=\frac{\pi}{\sqrt{2(E_3-E)}}\,,\nonumber
\end{eqnarray}
where $E_3$ is the ionization threshold energy of channel 3. The $1\to
2$ excitation probability is given by $P_{1;2}=|S^{phys}_{2;1}(E)|^2$
and is shown in the lowest panel of Fig.~1. The probability depends
strongly on the energy owing to the Rydberg series of resonances
attached to channel 3. We have derived an analytical expression for
the excitation probability $\bar P_{1;2}$ averaged over energy in this
region. Since $P_{1;2}$ is a periodic function of $\beta$, we average
over just one period $[0,\pi]$, i.e. over just one unit in the
effective quantum number of the closed channel 3:

\begin{eqnarray}
\bar P_{1;2}=\frac{1}{\pi}\int_0^{\pi}|S^{phys}_{2;1}(E)|^2d\beta\,.
\end{eqnarray}
We change this into a contour integral by making the variable change
\begin{eqnarray}
z={\rm e}^{2\rmi\beta}\,,dz={\rm e}^{2\rmi\beta}2\rmi d\beta\,,
\end{eqnarray}
and we obtain the following form for $\bar P_{1;2}$:
\begin{eqnarray}
\label{eq:int}
\frac{1}{2\pi\rmi}\oint_{unit\,
circle}\frac{dz}{z}\frac{\left[S_{2;1}(S_{3;3}-\frac{1}{z})-S_{2;3}S_{3;1}\right]
\left[S^*_{2;1}(S^*_{3;3}-z)-S^*_{2;3}S^*_{3;1}\right]}
{(S_{3;3}-\frac{1}{z})(S^*_{3;3}-z)}\,.\nonumber
\end{eqnarray}
This integral has two simple poles $z=0$ and $z=S^*_{3;3}$ inside the
circle. The residue of the integrand at $z=0$ is
\begin{eqnarray}
\frac{|S_{2;1}|^2S^*_{3;3}-S_{2;1}S^*_{3;1}S^*_{2;3}} {S^*_{3;3}}\,
\end{eqnarray}
and the residue at $z=S^*_{3;3}$ is
\begin{eqnarray}
\frac{S^*_{3;1}S^*_{2;3}\left(|S_{3;3}|^2S_{2;1}-S_{2;1}-S_{3;1}S_{2;3}S^*_{3;3}\right)} {S^*_{3;3}\left(|S_{3;3}|^2-1\right)}\,.
\end{eqnarray}
Once these are added, we see that the integral in Eq. (\ref{eq:int})
can be simplified to give the average excitation probability in the
energy range of rotational autoionizing states as
\begin{equation}
\label{eq:Pav}
\bar P_{1;2}= |S_{2; 1}|^2+ \frac{|S_{3; 1}|^2\ |S_{3; 2}|^2}{|S_{3;1}|^2+|S_{3; 2}|^2}\,.
\end{equation}

We can notice that the above formula is consistent with the fact that
Rydberg resonances are important when $|S_{1; 3}|^2$ is comparable in
magnitude with $|S_{1; 2}|^2$ and $|S_{2; 3}|^2$. Thus, the larger
$|S_{1; 3}|^2$ is, the larger are the resonances and their average
contribution to the total excitation cross section. It should be also
noted that the ratio
\begin{equation}
R=\frac{\bar P_{1;2}}{|S_{1; 2}|^2}
\end{equation}
is actually independent of the quantum defects and is entirely
controlled by Clebsch-Gordan algebra. In the case of the (1,
1)$\rightarrow$ (2, 1) transition, $R$ is equal to 15/7, leading to an
analytically averaged value of $\bar P_{1;2}=0.33$, to be compared
with the numerically-calculated average value of 0.29 (see
Fig.~1). The accuracy of the present analytic approach is therefore
estimated to be of the order of 10~\%.

Finally, we note that the present analytical averaging can of course
be applied to any SF $S$-matrices, in particular ANR $S$-matrices that
would have been converted to the SF frame of reference. In the
following, however, the averaging procedure will be implemented for
MQDT-RFT probabilities only.

\subsection{Detailed balance}

The principle of detailed balance, which is the consequence of the
invariance of the interaction under time reversal, states that the
transition probabilities $|{S_{1; 2}}|^2$ and $|{S_{2; 1}}|^2$ for a
certain inelastic process $|1\rangle\rightarrow |2\rangle$ and its
reverse $|2\rangle\rightarrow |1\rangle$ are equal {\it at a given
total energy, $E_{\rm tot}$}, defined as:
\begin{equation}
E_{\rm tot}=E_{k, 1} + E_1 = E_{k, 2} + E_2,
\end{equation}
where $E_{k, i}$ is the electron kinetic energy and $E_i$ is the
target internal energy. Following this definition, cross sections must
obey the formula:
\begin{equation}
\sigma_{1\rightarrow 2}(E_{k, 1})g_1E_{k, 1} = \sigma_{2\rightarrow
1}(E_{k, 2})g_2E_{k, 2},
\label{det1}
\end{equation}
where $g_i$ is the statistical weight of the state $i$. At the MQDT-RFT
level of theory, detailed balance is rigorously satisfied, even with
the above averaging of probabilities over Rydberg resonances. It
should be noted, however, that this would not be the case 
for the averaged formulae presented above, if cross
sections or rate coefficients were averaged over 
resonances instead of probabilities. At
the ANR level of theory, the assumption of target-state degeneracy
leads to transition probabilities that are equal {\it at a given
kinetic energy}. As a result, ANR cross sections obey the following
formula:
\begin{equation}
\sigma^{\rm ANR}_{1\rightarrow 2}(E_{k})g_1 = \sigma^{\rm ANR}_{2\rightarrow
1}(E_{k})g_2,
\label{det2}
\end{equation}
and Eq.~(\ref{det1}) is generally not satisfied at the ANR level. This
problem again reflects the ambiguity in defining the BF asymptotic
electron energy. In particular, if this energy is taken equal to the
entrance electron energy for an excitation process and to the exit
electron energy for the reverse deexcitation process, then
Eq.~(\ref{det1}) is rigorously satisfied at the ANR level. Moreover,
when the transition probability is energy independent, as it is at the
MQDT-RFT level for energies above the highest channel, then cross sections
obey both Eqs.~(\ref{det1}) and (\ref{det2}). This will be further
discussed in the next section.

\section{Results}

Rotationally inelastic cross sections are presented in Figs.~2 and 3
for upward rotational transitions in {\it ortho-} and {\it para-}
H$_3^+$. In Fig.~2, the rovibrational MQDT result is compared to the
MQDT-RFT and ANR calculations for the transition (1, 1)$\to $ (2,
1). Note that the full MQDT-rovibrational and MQDT-RFT results include
contributions from $J_{\rm tot}=1$ and $J_{\rm tot}=2$. We can first
notice that the agreement between MQDT-RFT and ANR cross sections is
extremely good above the highest (3, 1) channel, with relative
differences of less than 2~\%. In this energy range, we also notice in
the full MQDT treatment a number of narrow resonances that correspond
to vibrationally excited states of bound H$_3$ Rydberg states, as
observed in Fig.~1 for $J_{\rm tot}=2$. Between the (2, 1) and (3, 1)
channels, the ANR cross section is found to significantly
underestimate the averaged cross section owing to the appearance of
large purely rotational Rydberg resonances. In this energy range, the
MQDT-RFT calculation includes the analytical averaging of
probabilities over the Rydberg series and it provides, as expected, a
good averaged description of the full MQDT rovibrational result. We
also notice that despite a large effect of resonances on the
excitation probabilities for $J_{\rm tot}=2$ (a factor of 2, see
Fig.~1), the resulting effect on the cross sections summed over
$J_{\rm tot}$ is significantly lower, with an enhancement of about
30~\%. This of course was expected since the dominant contribution to
the cross section arises from a symmetry (here $J_{\rm tot}=1$) with
no purely rotational Rydberg series.

\begin{figure}
\begin{center}
  \rotatebox{-90}{\resizebox{90mm}{!}{
      \includegraphics*{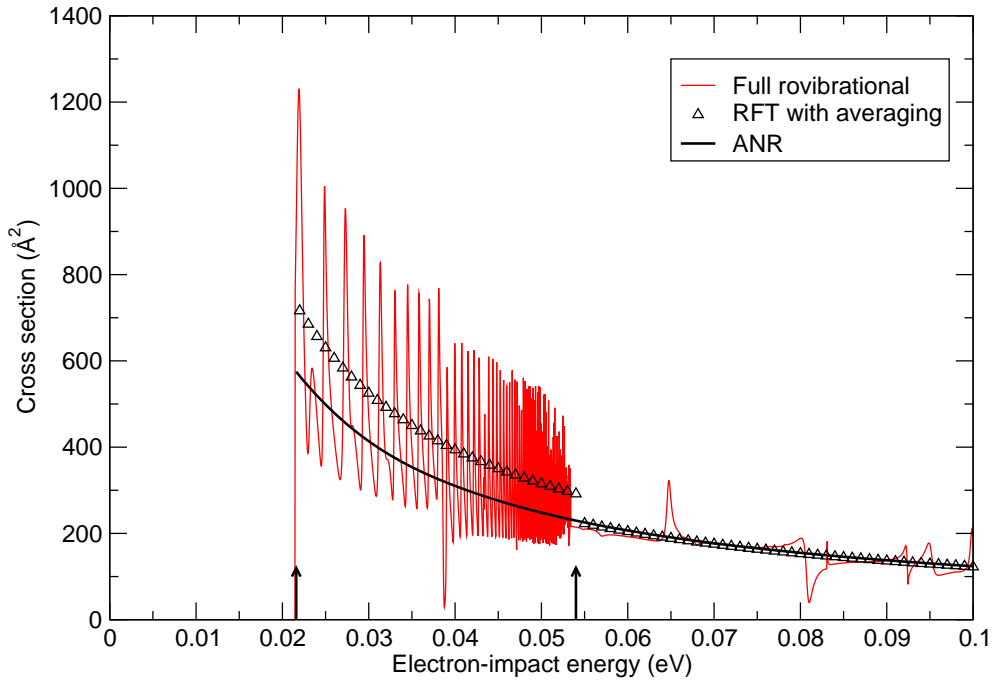}}}
\caption[]{Rotationally inelastic excitation cross sections for the transition
  (1, 1) $\rightarrow$ (2, 1) in H$_3^+$ as a function of
  electron-impact energy. The red line refers to the full
  rovibrational frame transformation calculation. The black line give
  the ANR cross section while triangles denote the MQDT-RFT cross
  section. Vertical arrows denote the (1, 1) $\to$ (2, 1) and (1, 1)
  $\to$ (3, 1) thresholds.}
\end{center}
\end{figure} 

Below the energy of the lowest ionization channel, the excitation
cross section in any MQDT treatment is identically zero, or strictly
speaking, the cross-section is not defined there because the electron
cannot be at infinity when all channels are closed. This behavior of
the cross-section agrees with Wigner's laws \cite{wigner48,stabler63}.

It is important to note that the ANR excitation cross section follows
the correct threshold law for an inelastic electron collision with a
positive ion:
\begin{equation}
\lim_{E'_k\to 0} \sigma^{\rm ANR}_{JK\to J'K'}(E_k) \propto
\frac{1}{E_k}=\mbox{non-zero constant}
\end{equation}
This actually reflects the weak energy dependence of the ANR
$T$-matrix elements, in agreement with the MQDT-RFT results (see
section~2.2). As explained in section~2.1, however, the ANR cross
sections are actually non-zero below threshold owing to our definition
of the ANR energy. In Figs.~2 and 3, cross sections were therefore
multiplied by the following Heaviside step function :
\begin{equation}
H(E_k) = \left\{
\begin{array}{lr}
1 & \mbox{if $E_k \geq E_{th}$} \\
0 & \mbox{if $E_k < E_{th}$},
\end{array}
\right.
\end{equation}
where $E_{th}=E_{J'K'}-E_{JK}$ is the threshold energy. This ad-hoc
correction is the simplest way of forcing the cross section to zero
below threshold while keeping a finite value at threshold, in
agreement with the above theoretical considerations. The effect of
this 'Heaviside correction' can be checked carefully in transitions
with no rotational Rydberg series, that is, those with $\Delta
J>1$. In Fig.~3, we can thus observe that all ANR cross sections for
transitions with $\Delta J=2$ are in excellent agreement with those
obtained at the MQDT-RFT level down to threshold. Of course, the exception
is the $(2, 1)\to (3, 1)$ transition in which closed-channels effects
occur at energies between the (3, 1) and (4, 1) channels and are of
similar magnitude as in $(1, 1)\to (2, 1)$. Note that we actually
observe a slight increase of closed channel effects as the initial $J$
increases. Rydberg resonances were thus found to enhance the $(4,
1)\to (5, 1)$ cross section by about 45~\% at energies between the (5,
1) and (6, 1) channels.

\begin{figure}
\begin{center}
  \rotatebox{-90}{\resizebox{90mm}{!}{
      \includegraphics*{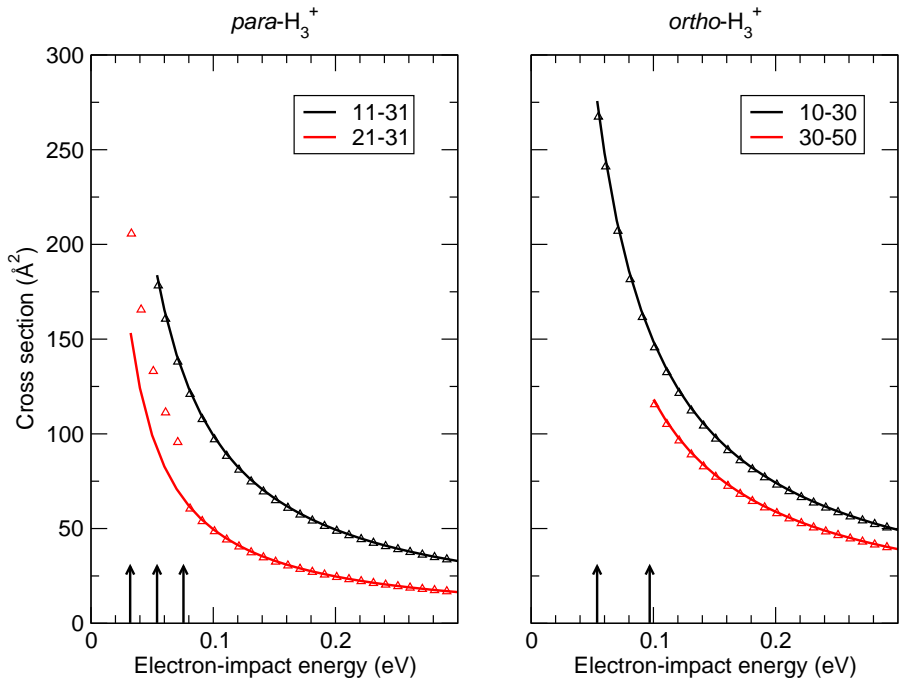}}}
\caption[]{Rotationally inelastic cross section for four different
  transitions in H$_3^+$ as a function of electron-impact
  energy. Lines give ANR cross sections while triangles denote MQDT-RFT
  cross sections. Vertical arrows denote the involved thresholds.}
\end{center}
\end{figure} 

The results presented in Figs.~2 and 3 show that despite the important
contribution of closed-channel effects in transitions with $\Delta
J=1$, the ANR theory is generally quite successful in predicting
rotational cross sections for molecular ions down to threshold. This
result was actually predicted by \citeasnoun{chang70}, although the
role of closed-channels was not considered by these authors. In their
study, \citeasnoun{chang70} showed that the correction term to the
adiabatic equation (Eq.~2.1 of their paper) is of the order of the
ratio of the electron mass $m$ to that of the nuclei $M$. They
concluded that the ANR theory should be accurate down to an energy:
\begin{equation}
E_k\gg(m/M)E_{th},
\end{equation}
that is, very close to threshold. It is interesting to notice that the
same authors estimated that the lower limit in the case of neutral
target molecules is much larger and is given by $E_k\geq
1.65E_{th}$. This huge difference reflects the influence of the strong
Coulomb field in the case of ions. 

We finally note that the above Heaviside correction does not of course
correct for the breakdown of detailed balance within ANR theory. As
discussed in section~2.3, a possible method to enforce detailed
balance is to use a different definition of the ANR energy for
excitation and deexcitation processes. However, as the present ANR
$T$-matrix elements are nearly energy independent, detailed balance is
actually almost satisfied at the ANR level. For practical applications
such as rate coefficient calculations, we still recommend computing
cross sections and rates for the excitation processes (including the
Heaviside correction) and using the detailed balance relation for the
reverse ones.

\section{Conclusions}

Near-threshold rotational excitation of H$_3^+$ by electron-impact has
been investigated by comparing ANR cross sections with calculations
based on the MQDT rotational-frame-transformation (MQDT-RFT) method
\cite{kokoouline03}. Very good agreement between ANR and MQDT-RFT
cross sections is obtained for kinetic energies above the resonance
regime caused by rotational closed-channels. These resonances occur
for transitions with $\Delta J=1$ and $\Delta K=0$ and for these, an
analytical formula for averaging transition probabilities over the
resonance structure has been formulated. This averaging procedure is
shown to significantly enhance probabilities and, to a lesser extent,
cross sections. In the case of transitions with $\Delta J>1$, the ANR
theory is shown to be accurate down to threshold, provided a simple
'Heaviside correction' is applied to the excitation cross sections. An
alternative but strictly equivalent solution is to interpret the ANR
energy as the exit kinetic electron energy and to compute deexcitation
cross sections.

Some previous studies of the rotational excitation of molecular ions
by electron collision, such as the H$_3^+$ study by
\citeasnoun{faure02c}, applied the ``correction'' factor $k'/k$,
designed for neutral targets by \citeasnoun{morrison88}, in a manner
that is not strictly correct.  The resulting error in those results
will be generally modest, nevertheless, except at energies very close
to rotational thresholds. Ideally, those results should be revised
along the lines implied by the more detailed analysis of the present
study, including the average effects of closed channel resonances
where appropriate. But this is beyond the scope of the present
article, so we do not present a comprehensive set of revised
excitation rate coefficients here. Moreover, the scattering matrices
obtained in $R$-matrix calculations can be used even below
closed-channel thresholds, provided they are sufficiently smooth
functions of energy, and provided they are used in combination with
MQDT closed-channel elimination formulas. The excellent agreement
between R-matrix ANR cross sections and the MQDT-RFT results in
Figs.~2 and 3 provides supporting evidence for this point. This is
also supported by the quantum defects as deduced from the $R$-matrix
calculations (using the diagonal BF $T$-matrix elements restricted to
the $p$-wave channels): we obtain ${\mu}_{0}$=0.05 and ${\mu}_{\pm
1}$=0.37, in very good agreement with the MQDT-RFT values of,
respectively, 0.05 and 0.39 (see section~2.2.1).

The range of validity of the adiabatic theory is therefore much wider
than the usual classically derived condition that the impacting
electron energy be large compared to the threshold energy. Moreover,
it should be stressed that H$_3^+$ is quite an unfavorable system for
the ANR theory owing to its large rotational spacings that make
threshold and closed-channel effects important up to large kinetic
energies ($>0.01$~eV). In particular, for astrophysical applications
where inelastic rate coefficients are required down to
$\sim$10$-$100~K \cite{faure06}, Rydberg resonances will play a
significant role. In contrast, for ions with rotational spacings
smaller than typically 10~K, resonances will be of minor
importance. Finally, in the case of strongly polar ions
($\mu\gtrsim$2~D) such as HCO$^+$, rotational excitation is completely
dominated by (dipolar) transitions with $\Delta J=1$ (see,
e.g. \citeasnoun{faure01}). In this case, resonances due to the low
probability transitions $\Delta J>1$ should not contribute
significantly to the dipolar cross sections. High-partial waves
($l>1$) are, however, required to converge dipolar cross sections
(i.e., the use of BF $T$-matrix elements actually leads to a divergent
cross section). It will be thus important to investigate in future
works the influence of closed-channels on $d$-, $f$- and higher
partial wave contributions to the cross section.

\ack
 
This work has been supported by the National Science Foundation under
Grant No. PHY-0427460 and Grant No. PHY-0427376, by an allocation of
NERSC supercomputing resources. AF acknowledges support by the CNRS
national program "Physique et Chimie du Milieu Interstellaire".

\end{document}